\documentclass[aps,prb,showpacs,twocolumn]{revtex4}%
\usepackage{graphicx}
\usepackage{color}
\usepackage{amsmath}
\usepackage{amsfonts}
\usepackage{amssymb}%
\newcommand{\be}{\begin{equation}}
\newcommand{\ee}{\end{equation}}
\newcommand{\bea}{\begin{eqnarray}}
\newcommand{\eea}{\end{eqnarray}}

\begin{document}

\title{Friedel oscillations induced surface magnetic anisotropy} 
\author{A. Szilva,$^1$ S. Gallego,$^{2}$
M.~C. Mu\~noz,$^{2}$ B.~L. Gy\"orffy,$^{3}$  G. Zar\'and,$^{1}$ and L. Szunyogh$^{1}$}
\affiliation{$^{1}$~Department of Theoretical Physics, 
Budapest University of Technology and Economics, Budafoki \'ut. 8, 1111
Budapest, Hungary \\
$^{2}$~Instituto de Ciencia de Materiales de Madrid, Consejo Superior de
Investigaciones Cient{\'{\i}}ficas, Cantoblanco, 28049 Madrid, Spain \\
$^{3}$~H.H. Wills Physics Laboratory, Bristol University, Royal Fort, Tyndall
Avenue, Bristol BS8 1TL, U.K.} 
\date{\today}

\begin{abstract}
We present detailed numerical studies 
of the magnetic anisotropy energy of a magnetic  impurity near the surface
of  metallic hosts (Au and Cu), that we describe in terms of
a realistic tight-binding surface Green's function technique.
We study the case when spin-orbit coupling originates
from the   {\it d}-band of the {\em host} material and 
we also investigate the case
of a  strong {\em local} spin-orbit coupling on the  impurity itself. 
The splitting of the impurity's 
spin-states is calculated to leading 
order in the exchange interaction between the impurity and the host atoms
using a diagrammatic Green's function technique.
The magnetic anisotropy constant is an oscillating function of the separation
$d$ from the surface: it asymptotically decays  as $\sim 1/d^{2}$ 
and its oscillation period   
is determined by the extremal vectors of the host's  Fermi Surface. 
Our results clearly show that the  host-induced magnetic anisotropy energy 
is by several orders of magnitude smaller than the anisotropy induced by the local
mechanism, which provides sufficiently large anisotropy values 
to explain the size dependence of the Kondo resistance observed experimentally.
\end{abstract}

\maketitle

\section{Introduction}
It is by now more than fifteen years ago that a surprising suppression of the Kondo effect in
thin films and wires of dilute magnetic alloys has been observed.\cite{exp1,exp2,exp3,exp4} 
A few years after the first experiments, \'Ujs\'aghy {\em et al.} proposed that 
the most likely explanation of the experimental observations is a 
spin-orbit coupling induced {\em magnetic anisotropy} in the vicinity of the surface of the
films: In the presence  of a surface, spin-orbit (SO) coupling   gives rise to 
a level splitting of the impurity spin, and thus blocks the spin-flip
processes responsible for the Kondo effect \cite{orsi1,orsi2,orsi3}.
Indeed, later experiments seemed to be in agreement with this simple
scenario and confirmed the predictions that follow from it.\cite{exp5} 
Fitting the experimental data for a $Au(Fe)$ film, 
\'Ujs\'aghy {\em et al.}\cite{orsi1}
estimated the width of the `dead layer', $L_c$,
where the splitting is larger or 
comparable to the Kondo temperature, $T_K = 0.3 K \simeq  
0.03$ meV, and obtained $L_c \simeq 180$~\AA .

To explain the  unexpectedly large width of the dead layer, 
\'Ujs\'aghy {\em et al.} also proposed a model to describe
surface-anisotropy,  which we shall refer to as the {\em host spin-orbit
coupling} (HSO) model. In this model 
an impurity with a half-filled $d$-shell and spin  $S=5/2$ is immersed in a host metal, 
where conduction  electrons  experience SO scattering 
through hybridizing with low-lying valence
$d$-orbitals of the host material.\cite{orsi1,orsi2,orsi3}  These calculations 
have been revised recently in Ref.~\onlinecite{orsi4}.
This HSO mechanism does not lead to the splitting of the  six-fold degenerate
spin state of the impurity, when placed in a bulk host 
with high (cubic or continuous rotational) symmetry.
However, the presence of the surface  induces an anisotropy term,
\be
H^{\rm HSO}_{\rm anis}=K(d) \left( {\bf n} \, {\bf S} \right)^2 \; ,
\label{H-anisHSO}
\ee
where ${\bf n}$ is the normal vector of the surface, $ {\bf S}$ is the spin-operator, and
$K(d)$ denotes the magnetic anisotropy constant at a distance $d$ from the 
surface. The anisotropy constant  $K(d)$ can be estimated in a simple free
electron model, by treating the spin-orbit coupling, $\xi$, and the exchange coupling, 
$J$, perturbatively. This calculation leads to the  asymptotic form,\cite{orsi4}
\be 
\label{K:orsi}
K(d) = A(k_F) J^2 \xi^2 \frac{\sin(2k_F d)}{d^3} \; ,
\ee 
where $k_F$ is the Fermi wavenumber.\cite{footnote1} 
Unfortunately, the constant $A(k_F)$
contains some cut-off parameters, which make the above formula less predictive
for the experiments.
However, ab initio calculations\cite{SG:prl97} indicated 
that this bulk mechanism is too weak to explain the experimental findings.

Recently, however, a rather different mechanism has been proposed to produce a 
magnetic anisotropy in the vicinity of a surface.\cite{SZG:prl06} 
This mechanism, which we shall refer to as {\em local spin-orbit
 coupling} (LSO) mechanism, 
assumes only a strong {\em local} SO coupling on the impurity's $d$-level. 
The basic observation leading to this mechanism is that, 
for partially filled $d$-shells, spin-states have also a large orbital
content. 
Therefore, spin states couple very strongly to
Friedel oscillations in the vicinity of a surface: electrons on the deep
$d$-levels can lower their energy by hybridizing
with the conduction electrons through virtual fluctuations. 
The corresponding anisotropy appears already to first order in the exchange
coupling $J$ and decays as $\sim 1/d^2$. In  the specific case  
of an impurity with a $d^1$ configuration, 
the corresponding $J_{3/2}$ ground state 
multiplet is split by the presence of the surface as\cite{SZG:prl06} 
\be
H^{\rm LSO}_{\rm anis}=K(d) \left( {\bf n} \, {\bf J} \right)^2 \;,
\label{H-anisLSO}
\ee
where ${\bf J}$ stands for the total angular momentum operators, and
$K(d) \sim J \sin(Q_F d)/d^{2}$,  with $Q_F$ being the length of an extremal vector of the
Fermi Surface (FS). As shown in Ref.~\onlinecite{SZG:prl06}
the anisotropy can  take the desired value of about a few tenths of meV even beyond 100 \AA \ 
from the surface.

Although the second (local) mechanism is expected to be dominant for
impurities with partially filled (not half-filled) $d$-shells, in
Ref.~\onlinecite{SZG:prl06} only a toy model, namely, a single-band metal
on a simple cubic lattice, has been considered.
For a quantitative comparison, however, and to decide, which mechanism is
responsible for the surface-induced anisotropy,  more realistic lattice and
band structures  should be used.
The aim of the present work is to provide such a qualitative and quantitative comparison of
the two mechanisms described above. For this purpose, we shall embed the 
impurity into an fcc lattice, and employ realistic tight-binding surface Green's function
methods\cite{sgfm}  to describe the conduction and valence electrons of the
host material. 
This method allows for a numerically exact treatment of the surface, and
also incorporates the  SO coupling non-perturbatively. 

To describe the magnetic impurity, we shall integrate out virtual charge
fluctuations on the $d$-level of the magnetic impurity, and construct 
realistic spin models, which take into account the specific magnetic and
crystal field structure of the impurity.\cite{CoxZawa}  

We shall then study the surface-induced anisotropy within both models, 
and  derive explicit expressions for the anisotropy constants in terms of the 
local density of states around the magnetic impurity.
Analyzing the behavior of $K(d)$ in the asymptotic regime, 
we find that the oscillations of $K(d)$ are related to the extremal vectors of
the Fermi surface.  For the case of Au and Cu host metals, we perform 
numerical calculations of the anisotropy 
constants as based on the asymptotic formulas and the oscillation periods 
are directly identified
from the numerically computed Fermi surface.  
In the case of the local SO model, we are also
able to confirm numerically the validity of the asymptotic expressions.
Our results support the priority of the local SO mechanism.

\section{Short review of the theoretical approach}

Before we present our results, let us to some extent 
outline the theoretical methods we use. 
As mentioned in the introduction, in  our approach we describe the
host material within  a tight binding Green's function formalism.
 The interaction between the magnetic impurity and the host, on the
other hand, is described in terms of an effective interaction, which 
we construct  by combining group theoretical methods with many-body
techniques. Once this effective exchange 
interaction Hamiltonian at hand, we can use relatively standard field
theoretical tools\cite{pseudofermion} to do perturbation theory in the exchange coupling, and
determine the surface-induced anisotropy.

\subsection{The Green's function of the host}

In this paper, we shall study  surfaces of ordered non-magnetic hosts, 
like the (001) surface of Au or Cu. In this case, 
the Hamiltonian of the host can be written as, 
\bea
\mathring{H}_{\lambda \sigma,\lambda^{\prime }\sigma^{\prime }}^{pn,p^{\prime }n^{\prime }} &=&
\left( \varepsilon _{\lambda }\,\delta _{\lambda \lambda ^{\prime }}
                                \delta _{\sigma \sigma^{\prime }} 
+ \xi \left( \mbox{\boldmath $\ell$} \cdot 
{\mathbf s} \right)_{\lambda \sigma,\lambda ^{\prime }\sigma^{\prime }}\right) \delta
_{pp^{\prime }}\delta _{nn^{\prime }}+
\nonumber \\
& +& V_{\lambda ,\lambda ^{\prime }}^{pn,p^{\prime }n^{\prime }}\,\delta _{\sigma \sigma^{\prime }}\quad,
\label{host-hamiltoni}
\eea
where $p$, $p^\prime$ denote atomic layers normal to the surface, 
$n$, $n^\prime$ label atomic sites within the layers, while $\lambda$, $\lambda^\prime$ 
denote the canonical $spd$ orbitals centered at the lattice positions and
$\sigma$, $\sigma^{\prime }$ are spin indices. 
In Eq.~(\ref{host-hamiltoni}), we replaced  all the parameters by their bulk
values,  i.e., we neglected the dependence of the on-site energies, 
$ \varepsilon _{\lambda }$ and the SO parameter, $\xi$, 
on the layer depth $p$. The hopping matrix elements, 
$V_{\lambda ,\lambda ^{\prime }}^{pn,p^{\prime }n^{\prime }}$ are confined to first
and next-nearest neighbors, and their layer-dependence is also neglected. 
These approximations  lead certainly to some errors in the calculated electronic
structure very close to the surface, however, they are expected to have no serious
 consequences in the asymptotic regime, which is the subject of our interest.
By the same token, in the vacuum  (i.e., $p \le 0$) the on-site 
energies are taken to be infinity. This simplifies somewhat our calculations, 
since only layers $p \ge 1$, forming thus a semi-infinite system,
need to be considered in the evaluation of the Green's function. 

The Hamiltonian Eq.~(\ref{host-hamiltoni}) can be recast into a matrix in the
spin and orbital labels,
\be
\underline{\mathring{H}}^{pn,p^{\prime }n^{\prime }} = \left\{ 
\mathring{H}_{\lambda \sigma,\lambda^{\prime }\sigma^{\prime }}^{pn,p^{\prime }n^{\prime }} \right\}
\; .
\ee
Since our system is translational invariant within the layers, we can also define the
Fourier transform of the Hamiltonian matrix, 
$\underline{\mathring{H}}^{pp^{\prime}}({\bf k})$, and introduce the 'semi-infinite' matrix,  
\be
\mathring{\cal H}({\bf k})=\left\{ \underline{\mathring{H}}^{pp^{\prime}}({\bf k}) \right\}_{p,p^\prime \ge 1} \; .
\ee
The resolvent or Green's function matrix is then given %in $k$-space 
as usual
\be
\mathring{\cal G}(z,{\bf k})=\left(  z-\mathring{\cal H}({\bf k}) \right)  ^{-1}\;,
\label{Gk}
\ee
with $z$ a complex number (energy).

To perform the matrix inversion in Eq.~(\ref{Gk}), we used 
the surface Green's function technique,\cite{sgfm} which is an efficient and, in principle, exact
tool to compute the Green's function. Most importantly, the computational
time of this method scales linearly with the number of layers,  for which the
Green's function is evaluated. The real-space representation
of the Green's function  can then be obtained by performing  the following Brillouin zone (BZ)
integral,
\be
\underline{\mathring{G}}^{pn,p^{\prime}n^\prime}(z)=\frac{1}{\Omega_{BZ}}\int d^{2}%
k\,\underline{\mathring{G}}^{pp^{\prime}}(z,{\bf k})\,e^{-i{\bf k}({\bf T}_{n^\prime}-{\bf T}_{n})}\;,
\label{GBZ}
\ee
where $\Omega_{BZ}$ is the volume of the 2-dimensional Brillouin zone, and the
translation vector ${\bf T}_{n}$ is related to the position of atom $n$ in
layer $p$ as ${\bf R}_{pn}= {\bf C}_{p} + {\bf T}_{n}$, with ${\bf C}_{p}$ a
layer-dependent reference position.

The host-Hamiltonian, Eq.\eqref {host-hamiltoni},
must be slightly modified 
in the presence of a magnetic impurity.  In this case,  the hopping of the 
conduction electrons to the impurity's $d$-orbitals 
should be excluded, 
since these  processes involve charge fluctuations 
at the magnetic impurity site, which 
will be incorporated in the  
effective exchange interaction (see next section). 
The simplest way to account for this constraint 
 is to shift the on-site $d$-state energies of the
impurity, $\varepsilon^i_\lambda$, far below the valence band, and add the
following term to the Hamiltonian, 
\be
\Delta H_{\lambda \sigma,\lambda^{\prime }\sigma^{\prime }}^{(q) pn,p^{\prime }n^{\prime }} 
= \left( \varepsilon^i_\lambda - \varepsilon_\lambda \right)
\delta_{pq} \, \delta_{p^\prime q} \, \delta_{n0} \, \delta_{n^\prime 0}\,  
\delta_{\lambda ,\lambda^{\prime }} \, \delta_{\sigma \sigma^\prime} \; ,
\label{deltaH}
\ee
where the impurity is at site $n=0$ and in layer $q$.

To describe the spin dynamics, we do not need to evaluate the full Green's
function: We  need its value only for 
 a small cluster of sites, ${\cal C}^{(q)}$, consisting of nearest
neighbor atoms  around the impurity and the impurity itself. 
Fortunately, since $\Delta H$ is also local, this restricted Green's function, 
\be
\underline{\underline{G}}^{(q)}(z) = 
\left\{ \underline{G}^{pn,p^{\prime}n^\prime} \right\}_{{\cal C}^{(q)}} \; ,
\ee
can be evaluated as 
\be
\underline{\underline{G}}^{(q)}(z) = \left( \underline{\underline{I}} - 
\underline{\underline{\mathring{G}}}^{(q)}(z) \, 
\Delta \underline{\underline{H}}^{(q)} \right)^{-1}
\underline{\underline{\mathring{G}}}^{(q)}(z) \; ,
\ee
where $\underline{\underline{I}}$ is the unit matrix, and the matrix elements of
$\underline{\underline{\mathring{G}}}^{(q)}(z)$ and 
$\Delta \underline{\underline{H}}^{(q)}$ are defined in 
Eqs.~(\ref{GBZ}) and (\ref{deltaH}), respectively.
Finally,  the spectral function matrix on this cluster 
is defined as
\be
\underline{\underline{\varrho}}^{(q)}
(\varepsilon)  =
-\frac{1}{2\pi i} \lim_{\delta \rightarrow +0} \left(  
\underline{\underline{G}}^{(q)}
(\varepsilon +i\delta) 
-\underline{\underline{G}}^{(q)}
(\varepsilon -i\delta) 
\right)  \;.
\label{romat}%
\ee
As shown in the following subsections, the 
 matrix elements of this spectral function matrix are directly related
to the magnetic anisotropy.

\subsection{Host spin-orbit  model of the magnetic anisotropy}

As in Refs.~\onlinecite{orsi1,orsi2,orsi3,orsi4} let us first consider a
spin $S=5/2$ impurity with a half-filled $d$-shell. 
In this case, we can neglect the SO interaction on the
magnetic ion, and the bulk SO interaction is the primary source
of the surface-induced anisotropy. 

\begin{widetext}
\begin{table}[h]
\centering
\begin{tabular}{lc}
\hline \hline
$|1\rangle$   & $D_{xz} = \frac{1}{2} \left( 
 s_{{\bf x} {\bf z}} + s_{\bar{\bf x} \bar{\bf z}}
-s_{{\bf x} \bar{\bf z}} - s_{\bar{\bf x} {\bf z}} \right)$            \\ 
$|2\rangle$   & $ D_{yz} = \frac{1}{2} \left( 
 s_{{\bf y} {\bf z}} + s_{\bar{\bf y} \bar{\bf z}}
-s_{{\bf y} \bar{\bf z}} - s_{\bar{\bf y} {\bf z}} \right)$            \\ 
$|3\rangle$   & 
 $D_{xy} = \frac{1}{2} \left( 
 s_{{\bf x} {\bf y}} + s_{\bar{\bf x} \bar{\bf y}}
-s_{{\bf x} \bar{\bf y}} - s_{\bar{\bf x} {\bf y}} \right)$           \\
$|4\rangle$   & $D_{x^2-y^2} = \frac{1}{2\sqrt{2}} \left( 
 s_{{\bf y} {\bf z}} + s_{\bar{\bf y} \bar{\bf z}}
+s_{{\bf y} \bar{\bf z}} + s_{\bar{\bf y} {\bf z}} 
-s_{{\bf x} {\bf z}} - s_{\bar{\bf x} \bar{\bf z}}
-s_{{\bf x} \bar{\bf z}} - s_{\bar{\bf x} {\bf z}} \right)$            \\ 
$|5\rangle$ \phantom{nnn}  & $D_{2z^2-x^2-y^2} = \frac{1}{2\sqrt{6}} \left( 
 2s_{{\bf x} {\bf y}} + 2s_{\bar{\bf x} \bar{\bf y}}
+2s_{{\bf x} \bar{\bf y}} + 2s_{\bar{\bf x} {\bf y}}
-s_{{\bf y} {\bf z}} - s_{\bar{\bf y} \bar{\bf z}}
-s_{{\bf y} \bar{\bf z}} - s_{\bar{\bf y} {\bf z}} 
-s_{{\bf x} {\bf z}} - s_{\bar{\bf x} \bar{\bf z}}
-s_{{\bf x} \bar{\bf z}} - s_{\bar{\bf x} {\bf z}} \right)$            \\ 
\hline \hline
\end{tabular}
\caption{Combinations of $s$-orbitals centered at the 12 neighbor sites around an impurity 
having the symmetry of atomic $d$ orbitals.}
\label{table:basis}
\end{table}
\end{widetext}

To construct the effective interaction between the host electrons and the
magnetic impurity, one can safely assume that the deep
$d$-levels of the magnetic impurity hybridize only with the $s$-orbitals of the 
neighboring host atoms. However, by symmetry, the deep $d$-levels can hybridize
only with appropriate linear combinations of these $s$-orbitals, 
$\alpha\in\{x^2-y^2,2z^2-x^2-y^2,xy,xz,yz\}$.
In case of an fcc lattice, e.g., we have  12 nearest neighbor $s$-orbitals,   
which 
can be labeled by  $s_{{\bf x} {\bf y}}$,  $s_{\bar{\bf x} \bar{\bf y}}$, 
$s_{{\bf x} \bar{\bf y}}$,  $s_{\bar{\bf x} {\bf y}}$, $\dots$,
$s_{{\bf y} \bar{\bf z}}$,  $s_{\bar{\bf y} {\bf z}}$, the subscripts
${\bf x} {\bf y}$ and $\bar{\bf x} \bar{\bf y}$ referring to neighboring sites at the positions
$a( \frac{\footnotesize{1}}{\footnotesize{2}}, \frac{\footnotesize{1}}{\footnotesize{2}},0)$
and
$a( -\frac{\footnotesize{1}}{\footnotesize{2}}, -\frac{\footnotesize{1}}{\footnotesize{2}},0)$
relative to the impurity, respectively, and $a$ denoting the cubic lattice 
constant. However, only 5 out of these 12 states
will have a $d$-wave character, and hybridize with the $d$-levels of the
magnetic impurity. These 5 states are listed in Table~\ref{table:basis}. 
Using these 5 spin-degenerate states, we can perform a 
Schrieffer-Wolff transformation\cite{schw} that leads to the following Hamiltonian, 
\begin{equation}
H_{J,ss^\prime}= \sum_{i=x,y,z} \sum_{\alpha=1}^5 J_\alpha 
\sum_{\sigma,\sigma^{\prime} = \pm 1}
c^\dagger_{\alpha\sigma} \, \sigma^i_{\sigma\sigma^{\prime}} c_ {\alpha \sigma^{\prime}}
\;   S^i_{ss^\prime}\;.
\label{H-HSO2}%
\end{equation}
Here $s,s^\prime=-\frac{\footnotesize{5}}{\footnotesize{2}},\dots,
\frac{\footnotesize{5}}{\footnotesize{2}}$ denote the $z$-components of the
impurity spins, $S^i$, and $\sigma^i$ denote the Pauli matrices. 
The operator $c^\dagger_{\alpha\sigma}$ creates a conduction electron with 
spin $\sigma$ in one of the states $|\alpha\rangle$ listed in Table~\ref{table:basis}.
In the bulk, only two of the exchange constants $J_\alpha$ are independent,
since by symmetry we have  $J_{xy} = J_{xz} = J_{yz}$ and $J_{x^2-y^2} =  J_{2 z^2-x^2-y^2}$.  
In the following, for the sake of simplicity, we shall set all these coupling
constants equal, and take $J_\alpha=J$. This
assumption does not modify our  conclusions.

The anisotropy induced by the surface can be computed by representing the spin 
in terms of Abrikosov pseudofermions, and then doing second order calculation
in the exchange coupling.\cite{orsi1} 
The zero temperature first and second order contributions to the
static ($\omega=0$) self-energy of the impurity spin can be expressed in terms
of the local density of states (spectral function) matrix, 
$\rho_{\alpha,\sigma;\alpha'\sigma'}$ as\cite{SZG:prl06}
\begin{align}
\Sigma_{s\,s^{\prime}}^{(1)}  &  =\int_{-\infty}^{\varepsilon_{F}}d\varepsilon
\,\mathrm{Tr}\left\{  \varrho(\varepsilon)\,H_{J,s\,s^{\prime}}\right\}
\nonumber\\
&  =J\sum_{i} \;S_{s\,s^{\prime}}^{i}
\int_{-\infty}^{\varepsilon_{F}}d\varepsilon\;
\mathrm{Tr}\left\{  \varrho(\varepsilon)\sigma^{i}\right\}  \; ,
\label{sigma1-HSO}
\end{align}
and
\begin{align}
\Sigma_{s\,s^{\prime}}^{(2)}  &  =
\int_{-\infty}^{\varepsilon_{F}} %d\varepsilon
\int_{\varepsilon_{F}}^{\infty} %d\varepsilon^{\prime}
\;\frac{d\varepsilon \, d\varepsilon^{\prime}}
       {\varepsilon^{\prime }-\varepsilon}
\sum_{\tilde{s}}\mathrm{Tr}\left\{  \varrho(\varepsilon
)\,H_{J,s\,\tilde{s}} \, \varrho(\varepsilon^{\prime})\,H_{J,\tilde{s}\,s^{\prime}%
}\right\} \nonumber\\
&  =J^{2}\sum_{i,j} \sum_{\tilde{s}} 
\;S_{s\,\tilde{s}}^{i} S_{\tilde{s}\,s^{\prime}}^{j} 
\int_{-\infty}^{\varepsilon_{F}} %d\varepsilon
\int_{\varepsilon_{F}}^{\infty} %d\varepsilon^{\prime}}^{j}
\;\frac{d\varepsilon \, d\varepsilon^{\prime}}
       {\varepsilon^{\prime }-\varepsilon}  \times
 \nonumber \\
& \qquad \qquad \qquad \qquad \qquad \mathrm{Tr}\left\{
\varrho(\varepsilon)\,\sigma^{i}\varrho(\varepsilon^{\prime})\,\sigma^{j}\right\}
\; ,   \label{sigma2-HSO}%
\end{align}
with Tr$\{\dots\}$ denoting the trace in the 10-dimensional subspace  
of the conduction electrons, and  $\varepsilon_{F}$  the Fermi energy.
The spectral function,  $\varrho_{\alpha,\sigma;\alpha'\sigma'}$, 
can easily be obtained from the real-space spectral function matrix elements,
  Eq.~(\ref{romat}).

Exploiting furthermore the tetragonal ($C_{4v}$) symmetry of an fcc(001) surface system and 
time-reversal invariance,
we find that $\varrho_{\alpha,\sigma;\alpha'\sigma'}$ has  the following
structure:
\be
\varrho=\left(
\begin{array}
[c]{ccccc}%
\varrho_{1}I_{2} & i\varrho_{5}\sigma_{z} & i\varrho_{6}\sigma_{x} & i\varrho_{7}\sigma_{y} & -i\varrho _{8}\sigma_{y} \\
-i\varrho_{5}\sigma_{z} & \varrho_{1}I_{2} & -i\varrho_{6}\sigma_{y} & 
 i\varrho_{7} \sigma_{x} & i\varrho_{8}\sigma_{x}\\
-i\varrho_{6}\sigma_{x} &  i\varrho_{6}\sigma_{y} & \varrho_{2}I_{2} & i\varrho_{9} \sigma_{z} & 0\\
-i\varrho_{7}\sigma_{y} & -i\varrho_{7}\sigma_{x} & -i\varrho_{9}\sigma_{z} & \varrho_{3}I_{2} & 0\\
 i\varrho_{8}\sigma_{y} & -i\varrho_{8}\sigma_{x} & 0 & 0 & \varrho_{4}I_{2}
\end{array}
\right) \label{rhoc4v} 
\ee
where $\forall \varrho_{i}\in\mathbb{R}$ and we dropped the energy argument of the spectral functions. 
The above form of $\varrho$ is fully confirmed by our  numerical calculations.
Inserting Eq.~(\ref{rhoc4v}) into Eqs.~(\ref{sigma1-HSO}) and
(\ref{sigma2-HSO}) yields,
$\Sigma_{s\,s^{\prime}}^{(1)}  \equiv 0$, and we find 
\be
\Sigma_{s\,s^{\prime}} \approx \Sigma^{(2)}_{s\,s^{\prime}} = C 
K_{\rm HSO} \, \left(S_{z}^{2}\right)_{s\,s^{\prime}} \; ,
\ee
where $C$ is a constant and the anisotropy constant, $K_{\rm HSO}$, can be expressed  as  
% C &  =2J^{2}\int_{-\infty}^{\varepsilon _{F}}d\varepsilon
%             \int_{\varepsilon _{F}}^{\infty }d\varepsilon^{\prime}
%\;\frac{1}{\varepsilon^{\prime}-\varepsilon}
%\left( 2\varrho_{1}(\varepsilon)\varrho_{1}(\varepsilon^{\prime})+
%  \varrho_{2}(\varepsilon)\varrho_{2}(\varepsilon^{\prime})+
%   \varrho_{3}(\varepsilon)\varrho_{3}(\varepsilon^{\prime})+
%   \varrho_{4}(\varepsilon)\varrho_{4}(\varepsilon^{\prime}) \right.
%\nonumber\\ & \left. \qquad \qquad \qquad \qquad \qquad \qquad \quad
%  -\varrho_{9}(\varepsilon)\varrho_{9}(\varepsilon^{\prime})-
%   \varrho_{5}(\varepsilon)\varrho_{5}(\varepsilon^{\prime}) \right) \; ,
%\\ \nonumber\\
\be
K_{\rm HSO}=K_{\rm HSO}^{6}+K_{\rm HSO}^{7}+K_{\rm HSO}^{8}-K_{\rm HSO}^{5}-K_{\rm HSO}^{9} \; ,
\ee
with
\be
K^i_{\rm HSO}   =-4J^{2}
\int_{-\infty}^{\varepsilon _{F}} d\varepsilon
\int_{\varepsilon _{F}}^{\infty} d\varepsilon^{\prime}
\;\frac{ \varrho_{i}(\varepsilon)\varrho_{i}(\varepsilon^{\prime})
}{\varepsilon^{\prime }-\varepsilon}  
\; .\label{Kc4v}%
\ee

If the impurity is placed in the bulk, then cubic symmetry further implies that
\be
\begin{array}{c}
\varrho_1(\varepsilon)=\varrho_2(\varepsilon) \:, 
\varrho_3(\varepsilon)=\varrho_4(\varepsilon) \:, 
\varrho_6(\varepsilon)=-\varrho_5(\varepsilon) \:,  \\
\varrho_8(\varepsilon)=\sqrt{3}\varrho_7(\varepsilon) \:,  
\varrho_9(\varepsilon)=-2 \varrho_7(\varepsilon) \; ,
\end{array}
\label{rhocubic}
\ee
and we obtain  $K_{\rm HSO}=0$. 
Thus the anisotropy is indeed  generated by the surface,
which breaks the cubic symmetry of the crystal.

\subsection{The local spin-orbit coupling  model of the magnetic anisotropy}

As in Ref.~\onlinecite{SZG:prl06}, let us now  consider a
magnetic impurity in a $d^{1}$ configuration such as a $V^{4+}$ or $Ti^{3+}$
ion. In this case, according to Hund's third
rule, a strong local spin-orbit coupling will lead to a $J=3/2$ multiplet that
is separated from the $J=5/2$ multiplet typically by an energy of the order of $\sim 1$ eV.
In a cubic crystal field, this $J=3/2$ ground 
multiplet remains degenerate ($\Gamma_{8}$ double representation), implying
that  no magnetic anisotropy appears if the magnetic impurity is  in the
bulk. Anisotropy will, however, arise, once the impurity is placed to the
vicinity of a surface that breaks the cubic symmetry. 

To construct the exchange interaction between the conduction electrons 
and the magnetic impurity, we first notice that the impurity's $J=3/2$
multiplet  can hybridize only with those linear
combinations of neighboring $s$-states which transform
according to the same ($\Gamma_{8}$) representation. 
Such a four--dimensional $d$-type set can be constructed from the states in
Table~\ref{table:basis}, as  
\bea 
|s_{-3/2} \rangle &=& D_{x^2-y^2} \,|\!\! \downarrow\rangle \; , \label{sm32}\\
|s_{-1/2} \rangle &=& D_{2z^2-x^2-y^2} \,|\!\! \downarrow\rangle \; , \label{sm12}\\
|s_{1/2}  \rangle &=& D_{2z^2-x^2-y^2} \,|\!\! \uparrow\rangle \; , \label{s12}\\
|s_{3/2}  \rangle &=& - D_{x^2-y^2} \,|\!\! \uparrow\rangle \; . \label{s32} 
\eea

Assuming that the impurity--host interaction is mainly dominated by quantum
fluctuations to the (non--degenerate) $d^{0}$ state, in lowest order of the
hybridization, a Coqblin--Schrieffer transformation leads to the following
effective exchange interaction, \cite{csch,SZG:prl06}
\begin{equation}
H_{J}=J\sum_{m,m^{\prime}}s_{m}^{\dagger}s_{m^{\prime}} \: \mid\! \frac{
\mbox{\footnotesize{3}} }{ \mbox{\footnotesize{2}} } m^{\prime}\rangle
\langle\frac{ \mbox{\footnotesize{3}} }{ \mbox{\footnotesize{2}} } m \!
\mid\;,
\end{equation}
where $|\frac{3}{2}m\rangle$ stand for the four states of the $\Gamma_{8}$
impurity multiplet, and $s_{m}^{\dagger}$  are creation 
operators creating an electron in the host states (\ref{sm32})--(\ref{s32}).

Interestingly, due to the different orbital content 
of the impurity states $|\frac{3}{2}, \pm \frac{3}{2}\rangle$ and
$|\frac{3}{2}, \pm \frac{1}{2}\rangle$, 
already the first order contribution to the self-energy gives a non-vanishing 
anisotropy in the vicinity of a surface,\cite{SZG:prl06}
\be
\Sigma_{mm^{\prime}}^{(1)}=J\int_{-\infty}^{\varepsilon_{F}}d\varepsilon\varrho
_{mm^{\prime}}(\varepsilon)\;.
\label{Sigma1-LSO}
\ee
The local spectral function of the host is now a $4\times 4$ matrix,
$\varrho_{mm^{\prime}}(\varepsilon)$, that has a diagonal structure, 
and is related to the spectral functions defined in 
Eq.~(\ref{rhoc4v}) as follows,
\bea
\varrho_{mm^{\prime}}(\varepsilon) &=& \varrho_{m}(\varepsilon) \, \delta_{mm^{\prime}} \; ,
\\
\varrho_{\pm 3/2}(\varepsilon)  &\equiv& \varrho_{3}(\varepsilon)  \: ,\phantom{nn} 
\varrho_{\pm 1/2}(\varepsilon)  \equiv \varrho_{4}(\varepsilon)  \: . 
\eea
From Eq.~(\ref{rhocubic}) it is obvious that the $J=3/2$ multiplet is degenerate 
under cubic symmetry (in the bulk), while under tetragonal symmetry it 
is split by an effective anisotropy term, Eq.~(\ref{H-anisLSO}),
with 
\be
K_{\rm LSO} = K^3_{\rm LSO} - K^4_{\rm LSO} \; ,
\ee
and
\be
K^i_{\rm LSO} = \frac{J}{2} \int_{-\infty}^{\varepsilon_{F}}d\varepsilon
\varrho_i(\varepsilon) \;.
\label{K-LSO}
\ee

\subsection{Asymptotic form of the anisotropy constants}

The presence of the surface induces Friedel oscillations 
in the local spectral functions.\cite{lang-kohn}
 For large distances $d$ from the surface, 
an asymptotic analysis can be performed based 
on the  rapid oscillations of the electronic wave function, $\sim e^{ik_z d}$.
In the simple case, when the constant energy surface in the three-dimensional
Brillouin zone of the bulk system
is formed by a single band (like the Fermi surface of noble metals),
this  leads to the following  expressions
for the spectral functions appearing in Eq.~(\ref{rhoc4v}),
\be
\varrho _{i}\left( \varepsilon,d \right) \simeq 
\varrho _{i}^{0}\left( \varepsilon \right) +\frac{1}{d}
\sum_n g^n_{i}\left( \varepsilon \right)
                  \cos\left[Q_{n}\left( \varepsilon \right) d + 
                    \theta_i^n\left( \varepsilon \right) \right]  \; ,
\label{asyrho}
\ee
where $\varrho _{i}^{0}\left( \varepsilon \right)$ is the spectral
function in the bulk, and the $Q_{n}\left( \varepsilon \right)$'s denote
the lengths of extremal vectors of the constant energy surface,
 normal to the geometrical surface. 
The factors
$g^n_{i}\left( \varepsilon \right)$ denote the amplitudes of the oscillations,
and $\theta^n_i\left( \varepsilon \right)$
are their phase. 
As we shall discuss later, in case of 
an fcc(001) geometry there are two different extremal vectors. 
Furthermore, it turns out that
each of the spectral function matrixelements has a non-negligible contribution
related only to one of these vectors,
therefore, as what follows, we shall label the extremal vectors 
with the index of the matrixelements $i$.
By substituting expression \eqref{asyrho}
into Eqs.~(\ref{Kc4v}) and \eqref{K-LSO} we then obtain the asymptotic form 
of the anisotropy constants.

\subsubsection{Host spin-orbit coupling model}

In case of the host spin-orbit coupling  model, 
the contributions, $K^i_{\rm HSO}$, to the magnetic anisotropy constant
can be expressed in leading order of $1/d$ as
\bea
K_{\rm HSO}^{i}&=&-\frac{4J^{2}}{d} \operatorname{Re} \int_{0}^{\infty } 
\frac{d \tilde{\varepsilon}}{\tilde{\varepsilon}}\, \label{Ki2} \\
&& \left\{
\int_{\varepsilon_{F}-\tilde{\varepsilon}}^{\varepsilon_{F}} d\varepsilon
\, \varrho_{i}^{0}\left(  \varepsilon + \tilde{\varepsilon} \right)
g_{i}\left(  \varepsilon\right)  
e^{i \left[ Q_i\left(  \varepsilon \right) d 
+ \theta_i\left( \varepsilon \right) \right]} \right. \nonumber \\
&& + \left.
\int_{\varepsilon_{F}}^{\varepsilon_{F}+\tilde{\varepsilon}} d\varepsilon
\, \varrho_{i}^{0}\left(  \varepsilon - \tilde{\varepsilon} \right)
g_{i}\left(  \varepsilon\right)  
e^{i \left[ Q_i\left(  \varepsilon \right) d 
+ \theta_i\left( \varepsilon \right) \right] } \right\} \; . \nonumber
\eea
Assuming that $\rho^{0}_{i}\left(  \varepsilon\right)$, 
$Q_{i}\left(  \varepsilon\right)$, 
$g_{i}\left( \varepsilon\right)$ and 
$\theta_{i}\left(  \varepsilon\right)$ are
slowly varying functions of $\varepsilon$, whereas
$e^{iQ_{i}\left(  \varepsilon \right) d}$ is rapidly oscillating,
the inner integrals in Eq.~(\ref{Ki2}) give sizable contributions 
only for small values of $\tilde{\varepsilon}$, 
and therefore, we can expand $Q_{i}\left( \varepsilon\right)$ 
around $\varepsilon _{F}$, $Q_{i}\left( \varepsilon\right) \simeq
Q_{i}\left( \varepsilon_F \right) +
Q_i^\prime\left( \varepsilon_F \right) 
\left(\varepsilon-\varepsilon_F\right)$ 
and substitute all the other functions by
their values at $\varepsilon_{F}$. This procedure yields the following 
asymptotic form: 
\be
K_{\rm HSO}^{i}=- \frac{4J^{2} \pi \varrho_{i}^{0}\left(  \varepsilon_F \right)
 g_{i}\left(  \varepsilon_F \right)}
{|Q^\prime_{i}\left(  \varepsilon_F \right)|}
\frac{\cos \left[ Q_{i}\left( \varepsilon_F \right) d 
+ \theta_i\left( \varepsilon_F \right) \right] }
{d^2}  
\; . \label{Kasy-HSO}
\ee

For free electrons, $Q\left( \varepsilon_F \right)=2k_F$, and  the
above result resembles that of \'Ujs\'aghy {\em et al.},\cite{orsi4}
however with a $\sim 1/d^2$  rather than a $\sim 1/d^3$ decay. This difference 
is a consequence of the assumption made in Ref.~\onlinecite{orsi4} that
 the scatterers in the host  are distributed homogeneously.

\subsubsection{Local spin-orbit coupling model}

In case of the local spin-orbit coupling  model
the energy integral in Eq.~(\ref{K-LSO}) can be easily performed yielding
\be
K^i_{\rm LSO} \approx 
\frac{J \, g_i\left(  \varepsilon_{F}\right) }
{2 |Q_i^{\prime}(\varepsilon_{F})|} 
\frac{\sin \left[  Q_i\left(  \varepsilon_{F}\right) d+
                   \theta_i \left(  \varepsilon_F \right) 
           \right] }{d^2} \;. 
\label{Kasy-LSO}
\ee
Interestingly, the asymptotic $d$-dependence of the magnetic anisotropy 
is described by very similar functions within both models, only the 
coefficients and the prefactors are different.

\section{Computational details}

For a realistic description of the host's valence and conduction bands
we used the on-site energies and the first and second nearest neighbor hopping
 parameters as given in Ref.~\onlinecite{papac} for Au and in 
Ref.~\onlinecite{cupar}  for Cu, and  set the cubic lattice constants to their 
experimental value cubic, $a_{\rm Cu}=3.615$ \AA \ and
$a_{\rm Au}=4.078$ \AA .\cite{webelements}
The spin-orbit parameter, $\xi$, has been determined from the difference 
of the SO-split $d$-resonance energies, $\Delta E_d= E_{j=5/2}-E_{j=3/2}$,
derived from self-consistent relativistic first-principles 
calculations.\cite{RSKKR} This splitting is related to our SO coupling as
\be
\Delta E_d \simeq \frac{5}{2} \xi  \;.
\ee
For Au we thus obtained $\xi= 0.64$ eV, while for Cu $\xi= 0.1$ eV. 
In order to reduce the computational efforts in performing the Brillouin zone 
integrals, Eq.~(\ref{GBZ}), we made use of the $C_{4v}$ point-group 
symmetry of the fcc(001) surface and applied an adaptive uniform mesh 
refinement for sampling the
$k$-points in the irreducible (1/8) segment of the Brillouin zone.
In general, about $10^4$ $k$-points were sufficient to calculate all the spectral 
function matrix elements in (\ref{rhoc4v}) with a relative accuracy of 1~\%.
We performed calculations for the $\varrho_i$'s for up to 50 monolayers 
below the surface, corresponding to a separation of $d\simeq90$ \AA \ and $d\simeq100$ \AA \
for Cu and Au, respectively.

Performing the double energy integral in Eq.~(\ref{Kc4v}) is a quite  
demanding numerical procedure. Therefore, for the host spin-orbit  model,  
we first fitted the spectral function matrix elements by the 
function \eqref{asyrho}, and then  used the asymptotic form, 
Eq.~(\ref{Kasy-HSO}) to compute the magnetic anisotropy, $K_{\rm HSO}$.
As we shall see later, beyond about 10 atomic layers
($d > 20$ \AA \ ) the calculated matrix elements followed the asymptotic
form, and the parameters, $g_i(\varepsilon)$, $\theta_i(\varepsilon)$
and $Q_i(\varepsilon)$ could be fitted with a high accuracy.

In case of the local spin-orbit coupling model, we also performed a similar
procedure to calculate the magnetic anisotropy constant in the asymptotic regime, 
Eq.~(\ref{Kasy-LSO}). However, in this case, it was also possible to 
compute the anisotropy constant directly  from Eq.~(\ref{K-LSO}): In this
case, we could deform the energy integration contour
by  using  the analyticity of the Green's function  
on the complex plane, and as few as 12 energy points along
a semi-circular contour in the upper complex half-plane were sufficient
for a very accurate evaluation of the corresponding integral.

\section{Results}

\subsection{Electronic structure of the bulk host}

\begin{figure}[ht!]
\includegraphics[
width=7cm,bb=10 10 210 260,clip]{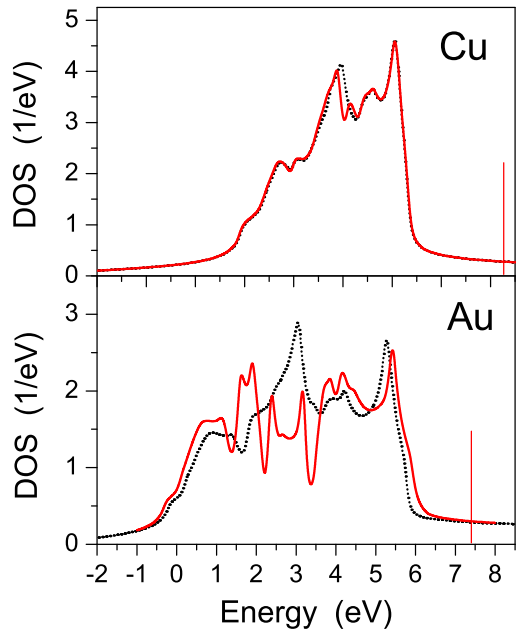}
\vskip -0.3cm
\caption{(Color online) 
Calculated valence band densities of states for Cu and Au bulk without SO interaction 
(dots) and with SO interaction (solid line). 
For the latter case the Fermi energies, 
$\varepsilon_{F}^{Cu}=8.3$~eV and $\varepsilon_{F}^{Au}=7.4$~eV, are 
indicated by vertical lines.
}
\label{fig:DOS}
\end{figure}

We first performed calculations of the density of states (DOS) of bulk Cu and Au.  
As shown  in Fig.~\ref{fig:DOS}, the dispersion of the $3d$-band of Cu is about 4 eV, while 
the $5d$-band of Au is much broader ($\sim$ 7 eV). Reassuringly, the positions and
the heights of the characteristic peaks of the DOS compare well with those obtained from self-consistent
first principles calculations.\cite{RSKKR,SKKR} Clearly, in copper, the small SO coupling,
$\xi=0.1$~eV, causes merely a slight modification in the DOS in the vicinity of the
$d$-like on-site energy ($\sim$ 5.07 eV). In the case of Au the SO coupling is
much stronger, $\xi=0.64$~eV, 
and is  large enough to influence the whole $d$-band: It gives  rise to  strong splittings of
the dispersion peaks and it also increases slightly  the bandwidth. As
indicated by the vertical   
lines in  Fig.~\ref{fig:DOS}, the Fermi energies,  
$\varepsilon_{F}^{Cu}=8.3$~eV and $\varepsilon_{F}^{Au}=7.4$~eV,
 lie well above the $d$-band for both metals.
\begin{figure}[ht!]
\includegraphics[
width=7cm,bb=10 10 220 260,clip]{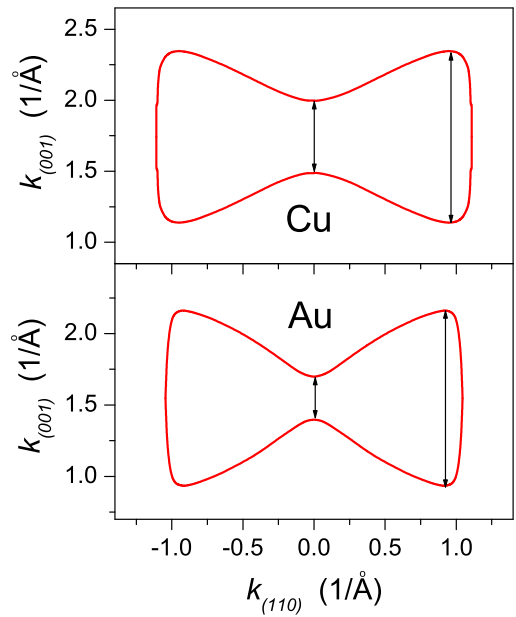}
\vskip -0.3cm
\caption{(Color online)
Calculated plane cuts perpendicular to the (1~-1~0) direction of the FS 
of Cu and Au. 
The arrows denote the extremal vectors of lengths,
$Q_{\rm min}^{\rm Cu}=0.505$~\AA$^{-1}$, $Q_{\rm max}^{\rm Cu}=1.208$~\AA$^{-1}$  and
$Q_{\rm min}^{\rm Au}=0.298$~\AA$^{-1}$, $Q_{\rm max}^{\rm Au}=1.228$~\AA$^{-1}$.
\label{fig:FS}}
\end{figure}

As we learned from  the asymptotic analysis presented 
in  Sec.~II.D,  extremal vectors of the Fermi surface
play a crucial  role  in determining the magnetic anisotropy constants. 
Therefore, we next investigated the plane cuts of the Fermi surface  perpendicular
to the (1~-1~0) direction. One can easily read off the 
length of the (001) extremal vectors from the
 cuts depicted in Fig.~\ref{fig:FS}: The absolute
minimum of the width of the Fermi surface, $Q_{\rm min}$, can be found at 
${\bf k}=0$, 
while the maximum width of the corresponding cut, $Q_{\rm max}$, is related 
to saddle-points of the Fermi surface. In the case of a Cu host the values obtained from
our tight-binding analysis,  
$Q_{\rm min}^{\rm Cu}=0.505$~\AA $^{-1}$ and 
$Q_{\rm max}^{\rm Cu}=1.208$~\AA $^{-1}$ correspond
to periods of 12.44 \AA \ and 5.20  \AA \ (6.88 and 2.88 monolayers (ML)) of
the oscillations, and 
agree fairly well with the periods, 6.08 ML and 2.60 ML,
 calculated by Lathiotakis {\em et al.}~\cite{iec-lath}. 
Similar satisfactory agreement can be found in the case of a Au host
between the periods found from our present calculations,
10.34 ML and  2.51 ML, and those calculated by Bruno and Chappert, 
8.6 ML and 2.6 ML.~\cite{iec-bruno}
It should be noted, however, that the shape of the FS depends very sensitively 
on the position of the Fermi energy the precise determination of which is a quite 
subtle task, since  for noble metals like Cu and Au the Fermi energy 
 lies in the very flat 4$sp$ band (see
also Fig.~\ref{fig:DOS}).
 
\subsection{The magnetic anisotropy constants within the host spin-orbit coupling model}

\begin{figure}[ht!]
\includegraphics[
width=7cm,bb=10 10 225 285,clip]{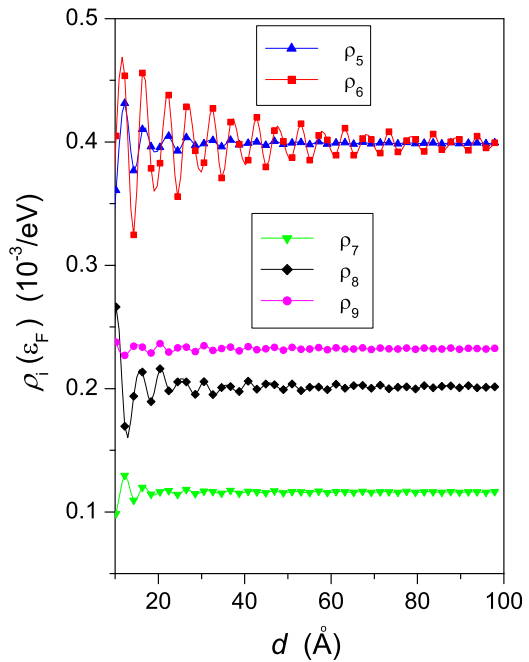}
\vskip -0.3cm
\caption{(Color online) Calculated off-diagonal spectral function matrix
  elements (see Eq.~(\ref{rhoc4v})),
at the Fermi energy as a function of the distance, $d$, 
from the (001) surface of a Au host. %Solid lines serve as guide for the eye.
\label{fig:rho-off}}
\end{figure}

We calculated the spectral function matrices, Eq.~(\ref{rhoc4v}),
at the Fermi energy of Cu and Au using the methods described
in Sections II.B. and C., for  up to 50 ML below the surface. As a convincing check of
our numerical procedure we verified that the
structure of the calculated matrices agrees with that derived analytically from 
symmetry principles. In the case of a Au host, in Fig.~\ref{fig:rho-off} we plotted the 
calculated off-diagonal matrix elements,
$\varrho_5(\varepsilon_F)$, $\dots$, $\varrho_9(\varepsilon_F)$, as a function of the
distance $d$ from the surface. As expected, large oscillations 
can be observed for all the spectral functions near the surface 
($d < 20$ \AA ).  These oscillations, however, survive for large
distances only for $\varrho_6$, while they are strongly damped in all the
other cases. The limiting values of $\varrho_i$ correspond to the bulk case and, as we checked, satisfy the 
conditions, Eq.~(\ref{rhocubic}) with less than 1\% relative numerical accuracy.

\begin{figure}[ht!]
\includegraphics[
width=7cm,bb=10 10 225 220,clip]{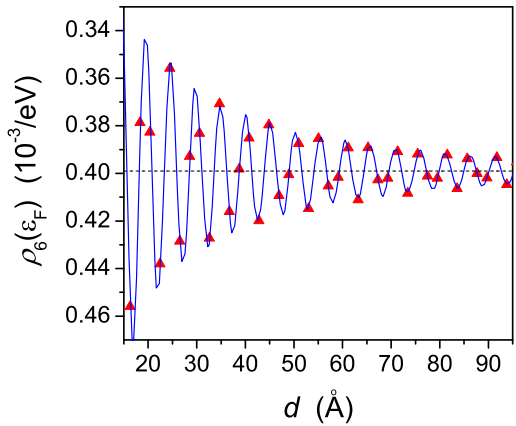}
\vskip -0.3cm
\caption{(Color online) Asymptotic fit to the function~(\ref{asyrho}) (solid line) of the 
calculated values of the $\varrho_6(\varepsilon_F)$ spectral function (triangles) as a
function of the distance from a Au(001) surface.
The dashed line denotes the bulk value of $\varrho_6(\varepsilon_F)$. 
\label{fig:rho6fit}}
\end{figure}

In Fig.~\ref{fig:rho6fit} we display the spectral function $\varrho_6(\varepsilon_F)$ 
on an enlarged scale, together with a fitting function of the 
form, Eq.~(\ref{asyrho}). Quite surprisingly, the asymptotic function 
	applies even in the range of $d \gtrsim 20$ \AA \ and, therefore, 
there is no need to perform a 'preasymptotic' analysis as suggested  in 
Ref.~\onlinecite{orsi4}. The fitted parameters of Eq.~(\ref{asyrho}) are as follows: 
$\varrho_{6}^{0}(\varepsilon_{F})$ = -3.99 $\pm$ 0.01 $\cdot$ $10^{-4}$ eV$^{-1}$,
$g_6(\varepsilon _{F})$ = -1.484 $\pm$ 0.008 $\cdot$ 10$^{-3}$ \AA eV$^{-1}$,
$Q_6(\epsilon _{F})$ = 1.2228 $\pm$ 0.0001 \AA $^{-1}$,
and $\theta_6(\epsilon_{F})$ = 1.324 $\pm$ 0.006 rad.
It is particularly noteworthy that the fitted wavenumber agrees with an accuracy of 0.5 \%
with the length of the extremal vector, $Q_{\rm max}$, computed from the Au Fermi Surface.
We could fit all other off-diagonal spectral function components 
entering the expression of $K_{\rm HSO}$
with a similar fit with exactly the same wavenumber. However, the amplitude
of these other components was  by at least two orders of magnitude smaller than $g_6(\varepsilon _{F})$.

\begin{figure}[htb]
\includegraphics[
width=7cm,bb=10 10 225 260,clip]{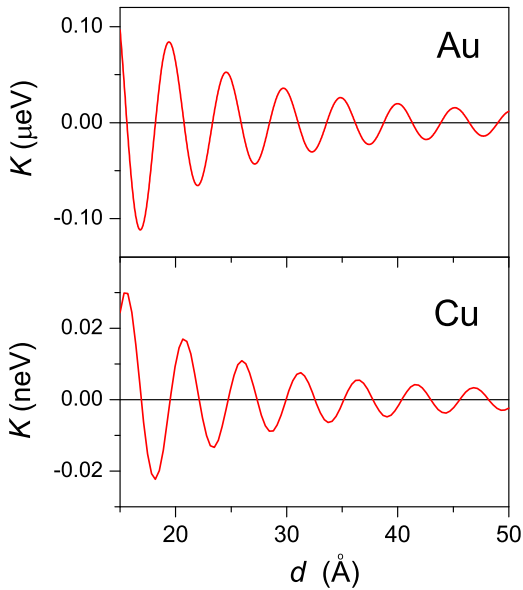}
\vskip -0.3cm
\caption{(Color online) Upper panel: The $K^{6}_{HSO}$ contribution 
to the magnetic anisotropy constant
within the host spin-orbit coupling model for a Au host as calculated from the asymptotic expression, Eq.~(\ref{Kasy-HSO}).
Lower panel: The $K^{9}_{HSO}$ contribution to the magnetic anisotropy constant in the case of a Cu host. 
In both cases an exchange interaction parameter $J=1$ eV was used.
\label{fig:khso}}
\end{figure}

Our calculations thus indicate that the long-wavelength oscillation corresponding to $Q_{\rm min}$
of the FS either enters with a negligibly small amplitude or 
doesn't enter at all in the asymptotic form of the off-diagonal spectral
function matrixelements. This can easily be understood by noticing that 
the asymptotic contributions  to
the real-space spectral function matrixelements, 
$\varrho_{s \sigma,s \sigma^\prime}^{(q+p)n,(q+p^\prime) n^\prime}(\varepsilon)$
($p,p^\prime=0,\pm1, d=q\frac{a}{2}$) related to $Q_{\rm min}$ are
of the following form,
\bea
\varrho_{s \sigma,s \sigma^\prime}^{(q+p)\,n,(q+p^\prime)\, n^\prime}(\varepsilon)
&\simeq& \varrho_{s \sigma,s \sigma^\prime}^{(0)p\,n,p^\prime \,n^\prime}(\varepsilon)
+ \frac{g_{s \sigma,s \sigma^\prime}^{p,p^\prime}(\varepsilon)}{d} 
\times \nonumber \\
&& \cos\left[ Q_{\rm min}(\varepsilon)d + \theta(\varepsilon) \right] \; ,
\label{rsrho-qmin}
\eea
where $\varrho_{s \sigma,s \sigma^\prime}^{(0)p\,n,p^\prime \,n^\prime}(\varepsilon)$
refer to the corresponding bulk matrixelements. Eq.~(\ref{rsrho-qmin}) 
implies that the oscillating part does not depend on the in-plane positions, 
$n$ and $n^\prime$, which is the consequence that the minimal extremal vector  
is at the ${\bf k}=0$ point of the 2D Brillouin zone.
As explained  in Sec.~II.B., the matrixelements in Eq.~(\ref{rhoc4v}) are 
linear combinations of the above
real-space matrix elements according to the states in Table~I.
Since the states $|\alpha\rangle$ ($\alpha=1,\dots,4$) are constructed 
as antisymmetric combinations of
neighboring $s$-orbitals in the same plane, $q+p$, or as a sum of such 
antisymmetric combinations, in their matrixelements
the asymptotic oscillatory part corresponding to $ Q_{\rm min}$ necessarily cancels.
As a consequence, only the spectral function 
$\varrho_4 \equiv \langle 5 | \varrho | 5 \rangle$ 
has asymptotic oscillations with wavenumber $Q_{\rm min}$, which, however,
does not give a contribution in the host SO model.

We calculated the magnetic anisotropy constant 
using the asymptotic fits of the spectral functions and
Eq.~(\ref{Kasy-HSO}). We numerically determined the energy 
derivative of the magnitude of the extremal vector, $Q^\prime(\varepsilon_F)$, by  
fitting the spectral functions at two  energy values 
close below and above $\varepsilon_F$ and obtained 
$Q^{\prime }(\epsilon _{F})= 0.235$ (\AA \ eV)$^{-1}$.
Thus,  in case of a Au host we get the following asymptotic function for $K^6_{HSO}(d)$ 
(displayed in the upper panel of Fig.~\ref{fig:khso}) 
\be
K^{6}_{HSO}(d) = \frac{31.66}{d^2} \cos\left[ 1.2228 \cdot d + 1.324 \right] \: \mu{\rm eV} \; ,
\ee
where $d$ is measured in \AA . 
Notice the surprisingly small magnitude of $K^6_{HSO}$: 
even at a distance of about $d=20$ \AA \ the amplitude of the above 
oscillating function is about  0.079 $\mu$eV.

We performed similar calculations for a Cu host. In Cu, the spectral functions 
show asymptotic oscillations with $Q(\varepsilon_F)=1.205$ \AA$^{-1}$ that
agrees 
within 0.3 \% with the length of the
extremal vector, $Q_{\rm max}$, of the Cu FS. In Cu, the $K^9_{HSO}$ contribution shown in
the lower panel of Fig.~\ref{fig:khso} dominates the magnetic anisotropy. 
This is in the range of 0.01 neV = 10$^{-11}$ eV, i.e., it is 
at least by three orders of magnitude smaller than the one found in case a
Au host. This decrease is mostly  due to the 
smaller SO interaction in Cu than in Au. As we checked by varying $\xi$ for
Au,  the spectral functions in Eq.~(\ref{rhoc4v}) scale linearly with $\xi$,
therefore, by Eq.~(\ref{Kc4v}) the magnetic anisotropy constant scales as
$\sim \xi^2$. 
This result clearly justifies
the approach of \'Ujs\'aghy {\em et al.}, who treated the SO interaction 
perturbatively.\cite{orsi1,orsi2,orsi3,orsi4}.

\subsection{The magnetic anisotropy constants within the local spin-orbit coupling model}

As pointed out in  Ref.~\onlinecite{SZG:prl06}, a mechanism based on a strong local SO
interaction of the impurity (local SO model) can give rise to a level splitting
that is orders of magnitude larger than the host-induced anisotropy.
To demonstrate this idea, in Ref.~\onlinecite{SZG:prl06} we studied the simple
but unrealistic case of a single-band metal on a simple cubic lattice. Here 
we extend the calculations of Ref.~\onlinecite{SZG:prl06} and perform
calculations for realistic host metals (Cu and Au). 

\begin{figure}[htb!]
\includegraphics[
width=7cm,bb=10 10 235 260,clip]{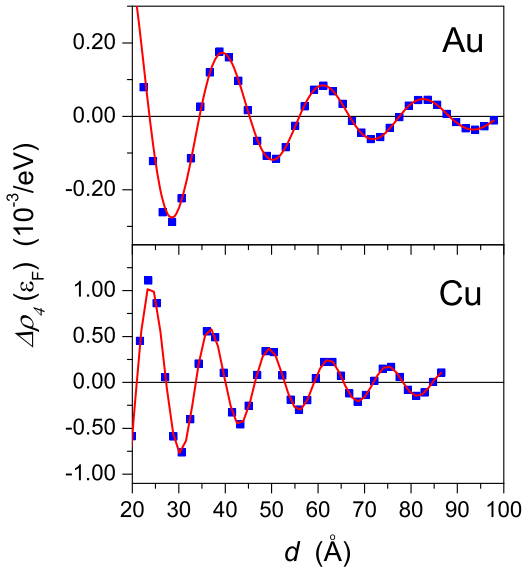}
\vskip -0.3cm
\caption{(Color online) Calculated values of $\Delta \varrho_4(\varepsilon_F) \equiv
\varrho_4(\varepsilon_F)-\varrho_4^{0}(\varepsilon_F)$ 
(squares) with corresponding asymptotic fits, Eq.~(\ref{asyrho}), (solid line) 
as a function of the distance from the (001) surface of Au and Cu.
\label{fig:deltarho}}
\end{figure}

According to the theory presented in Sec.~II.C, we need to compute the diagonal 
spectral function matrixelements,
$\varrho_3 \equiv \langle 4 | \varrho | 4 \rangle$ and 
$\varrho_4 \equiv \langle 5 | \varrho | 5 \rangle$, see Table~I
and Eq. (\ref{rhoc4v}).
Our calculations clearly showed that the $d$ dependent Friedel oscillations of $\varrho_3$ 
are several order smaller in magnitude than those of $\varrho_4$.
This can be understood by noticing that,
due to the different spatial character of these two states 
($D_{x^2-y^2}$ and $D_{2z^2-x^2-y^2}$),
$\varrho_3$ comprises an average of spectral weights in layers $q-1$ and $q+1$,
while $\varrho_4$ takes the difference of spectral weights in layer $q$ with respect
those in layers $q-1$ and $q+1$, $q$ denoting the layer of the impurity's position.
Recalling that for a cubic bulk $\varrho_3=\varrho_4$, 
see Eq.~(\ref{rhocubic}), in the asymptotic region 
$K_{\rm LSO}$ becomes proportional with the integral of 
the function, $\Delta \varrho_4(\varepsilon,d) \equiv
\varrho_4(\varepsilon,d) - \varrho_4^{0}(\varepsilon)$.  This function 
is displayed in Fig.~\ref{fig:deltarho} for both the Au and the Cu hosts.
Remarkably, the amplitude of the Friedel oscillations is about one
order of magnitude larger than those of the off-diagonal spectral functions 
(compare with Fig.~\ref{fig:rho-off} for the case of Au). Note that
the off-diagonal matrixelements appear 
in first order of the spin-orbit coupling. 
The oscillations have larger periods
as compared to the off-diagonal spectral functions: a fit to the asymptotic function, 
Eq.~(\ref{asyrho}), shown also in Fig.~\ref{fig:deltarho} gave the values
$Q^{Au}=0.292$ \AA$^{-1}$ and $Q^{Cu}=0.505$ \AA$^{-1}$, in very good 
agreement with the length of the small extremal vector, $Q_{\rm min}$, 
of the corresponding Fermi surfaces.
Interestingly, the amplitude of the
 oscillations is more than three times larger for Cu than for Au: 
from the fits we obtained 
$g_4(\varepsilon_{F})=1.16\cdot10^{-2}$ \AA \ eV$^{-1}$ and
$3.53\cdot10^{-2}$ \AA \ eV$^{-1}$ for the case of Au and Cu, respectively.
 
\begin{figure}[htb!]
\includegraphics[
width=7cm,bb=10 10 230 190,clip]{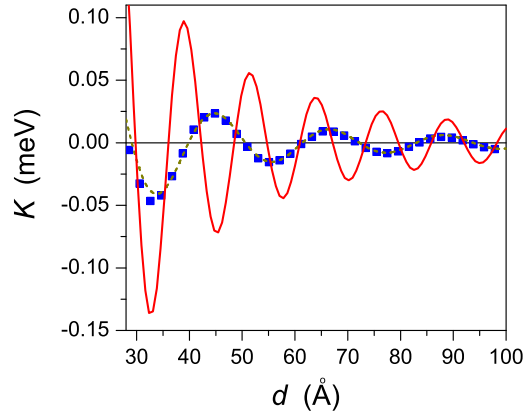}
\vskip -0.3cm
\caption{(Color online) Magnetic anisotropy constants within the local spin-orbit
coupling model calculated by using the asymptotic formula,
Eq.~(\ref{Kasy-LSO}), as a function of the distance $d$ from the (001) surface of a Au (dashes) 
and a Cu host (solid line). 
In case of Au the squares stand for the magnetic anisotropy constants calculated
directly from Eq.~(\ref{K-LSO}).
\label{fig:klso}}
\end{figure} 

Fig.~\ref{fig:klso} shows the magnetic anisotropy constants obtained using Eq.~(\ref{Kasy-LSO}) with
the parameters  extracted from the fits of $\Delta \varrho_4(\varepsilon _{F},d)$.
The parameter $Q^{\prime }(\epsilon _{F})$ was computed as for the off-diagonal spectral functions,
and took a value of  0.245 (\AA eV)$^{-1}$ for Au and 0.238 (\AA eV)$^{-1}$ for Cu.
Choosing again $J=1$ eV, we obtained for  the amplitudes of the oscillations of $K$, 
$A(d)=0.0237/d^2$ eV and $A(d)=0.0742/d^2$ eV ($d$ measured in \AA \ ) in Au and Cu, respectively.
In particular, for Cu this gives an amplitude of 0.03 meV at $d=50$ \AA  \
or 0.007 meV at $d=100$ \AA,
which is in the range of $T_K$ for typical dilute magnetic alloys, such as Cu(Mn), Cu(Cr).

In Fig.~\ref{fig:klso}, we also compare the magnetic anisotropy constants obtained 
from the asymptotic analysis with the values we get by performing the 
 contour integration in  Eq.~(\ref{K-LSO}). Apparently, already for $d > 35$
 \AA,  these values  lie almost perfectly on the asymptotic curve. This nice
 agreement  proves the validity of
the asymptotic formula, Eq.~(\ref{Kasy-LSO}), as well as the accuracy of our numerical procedure to
compute the magnetic anisotropy constant. 

\section{Summary and conclusions}

In this paper, we performed a theoretical 
study of two mechanisms for surface-induced magnetic anisotropy of a magnetic 
impurity: a local spin-orbit  mechanism (LSO),\cite{SZG:prl06}  and
a host spin-orbit mechanism (HSO).\cite{orsi1}  Both mechanisms appear as 
a result of Friedel-like oscillations in the local spectral functions, induced by the
surface. In the local SO mechanism, the rather large diagonal, i.e., {\em charge oscillations} 
couple through the local
spin-orbit coupling on the d- or f-level of the  magnetic impurity to the
impurity spin and lead to a surface-induced splitting of the spin states.
 The host SO mechanism, on the other hand, relies on 
oscillations in the {\em off-diagonal} elements of the local spectral functions, i.e., 
oscillations in the ``spin sector'' that couple directly to the spin through an exchange
interaction. These oscillations are induced
by the SO coupling in the host metal 
and, thus,  they are  much weaker than the Friedel oscillations in the ''charge sector''. 
Based upon this simple picture, one  therefore expects that the 
first mechanism is dominant for impurities with a partially filled d- or
f-shell, while the host SO mechanism may become important for half-filled 
shells, in which case the local SO mechanism cannot be at work.

In this paper we attempted to compare these two mechanisms quantitatively.
For the description of the host's valence and conduction electrons we used a 
tight-binding Green's function technique, which allows for a perfect treatment of the semi-infinite
surface geometry, and makes also possible a non-perturbative treatment
of the host SO interaction. We then used a field theoretical approach to 
compute the self-energy of the spin up to   
first (local SO) and second orders in the exchange coupling, $J$, (host SO
model), and derived explicit expressions for the anisotropy constants, $K$, 
as a function of the separation $d$ between  the impurity and the surface. 

These expressions have been analyzed using an asymptotic analysis which
resulted in a very similar oscillatory dependence of $K$ on $d$ in both models: 
the periods of the oscillations could be  identified as  the magnitudes of the extremal vectors of the
Fermi Surface of the bulk host and  their amplitudes decayed in both
models as 1/$d^2$. 
Here we must remark that in our calculations in Ref.~\onlinecite{SZG:prl06} 
we predicted a
$1/d^3$ decay of the oscillations of $K$  within the host spin-orbit
mechanism. This must be contrasted to the results of the present work, where
we find rather a $1/d^2$ scaling of the host-induced anisotropy.
This apparent controversy is due to a small difference in the calculations: 
Unlike the present work, in Ref.~\onlinecite{SZG:prl06} we neglected the potential scattering at the
impurity site, i.e., we used the local spectral functions of a perfect
semi-infinite host. In this case, however, one can show that  certain off-diagonal
elements of the local spectral function matrix must vanish 
due to two-dimensional translational symmetry. These off-diagonal
matrix elements are non-zero, once translational invariance is 
broken by potential scattering at the impurity site, 
and they give rise to a $1/d^2$ decay of the anisotropy as shown in Sec. II.D. 

Using realistic tight-binding parameters,
we calculated the amplitudes of the magnetic anisotropy oscillations
for the cases of Au and Cu metal hosts.
As expected from the very different SO interactions in these metals, 
within the host SO model, the
magnetic anisotropy constant for Au turned out to be about
three orders larger in magnitude than for Cu.
Nevertheless, even for a Au host and close to the surface,
the magnetic anisotropy constants remained below the range  of 0.1 $\mu$eV.
Though a direct comparison with the result of Ref.~\onlinecite{orsi4} is quite
questionable mainly due to the different geometrical distribution
of the host atoms and to the different approximations used,
the above value is close to the {\em lower} limit of the estimated range
of $K$ given in Ref.~\onlinecite{orsi4}.
We therefore conclude that most probably the host SO mechanism of Ref.~\onlinecite{orsi4}
is too weak to explain
the size-dependence of the Kondo resistance. 

The local SO mechanism proposed in Ref.~\onlinecite{SZG:prl06}, on the other hand, 
gives a magnetic  anisotropy constants for Cu  in the range of 
 0.03-0.01 meV for even at distances 50-100 \AA \ away  from the surface.
Although they are in the same range,  the magnetic  anisotropy constants
for Au were about three times smaller than the ones we got for Cu. 
Our numerical studies imply that the primary mechanism to produce a magnetic anisotropy in the
 vicinity of a surface is provided by the  local SO coupling, where 
the local  Hund's rule coupling conspires with Friedel oscillations to 
produce a large anisotropy effect. 
This mechanism seems to be  large enough to explain the suppression  of the Kondo resistance anomaly 
observed in thin films and it also
supposed to be the dominant source
of (random) magnetic anisotropy  in metallic  mesoscopic structures
such as metallic nano-grains, nano-wires, or point contacts.  

\bigskip
The authors are indebted to A. Zawadowski and  O. \'Ujs\'aghy for
valuable discussions. This work has been financed by the Hungarian National
Scientific Research Foundation (OTKA T068312 and 
NF061726) and
by a cooperation between the Spanish Ministry of Science and
the Hungarian Science and Technology Foundation
 (HH2006-0027 and OMFB-01230/2007).


\begin{thebibliography}{99}
\bibitem{footnote1}
We note that Eq.~(\ref{K:orsi}) refers to a homogeneous distribution of
host atoms (spin-orbit scatterers), while for the case of a layered
geometry and phase coherence, $k_F a = p \pi$, $a$ and $p$ 
being the interlayer distance and an integer, respectively,
a slower decay, $K(d)\sim 1/d^2$, has been predicted.\cite{orsi4}
\end{thebibliography}
\end{document}